\documentclass[aps,prd,twocolumn,superscriptaddress]{revtex4}
\usepackage{epsfig,epsf}
\usepackage{amsmath}
\usepackage{amsthm}
\usepackage{amsfonts}
\usepackage{amssymb}
\usepackage{dsfont}
\usepackage{multirow}
\usepackage{appendix}
\usepackage{slashed}
\usepackage[active]{srcltx}
\usepackage{psfrag}

\begin{document}

\title{Weak decays of the axial-vector tetraquark $T_{bb;\bar{u} \bar{d}%
}^{-} $ }
\date{\today}
\author{S.~S.~Agaev}
\affiliation{Institute for Physical Problems, Baku State University, Az--1148 Baku,
Azerbaijan}
\author{K.~Azizi}
\affiliation{Department of Physics, Do\v{g}u\c{s} University, Acibadem-Kadik\"{o}y, 34722
Istanbul, Turkey}
\author{B.~Barsbay}
\affiliation{Department of Physics, Kocaeli University, 41380 Izmit, Turkey}
\author{H.~Sundu}
\affiliation{Department of Physics, Kocaeli University, 41380 Izmit, Turkey}

\begin{abstract}
The weak decays of the axial-vector tetraquark $T_{bb;\bar{u} \bar{d}}^{-}$
to the scalar state $Z_{bc;\bar{u} \bar{d}}^{0}$ are investigated using the
QCD three-point sum rule approach. In order to explore the process $T_{bb;
\bar{u} \bar{d}}^{-} \to Z_{bc;\bar{u} \bar{d}}^{0}l \bar{\nu_l}$, we
recalculate the spectroscopic parameters of the tetraquark $T_{bb;\bar{u}
\bar{d}}^{-}$ and find the mass and coupling of the scalar four-quark
system $Z_{bc;\bar{u} \bar{d}}^{0}$, which are important ingredients of
calculations. The spectroscopic parameters of these tetraquarks are computed
in the framework of the QCD two-point sum rule method by taking into account
various condensates up to dimension ten. The mass of the $T_{bb;\bar{u} \bar{%
d}}^{-}$ state is found to be $m=(10035~\pm 260)~\mathrm{MeV}$, which
demonstrates that it is stable against the strong and electromagnetic
decays. The full width $\Gamma$ and mean lifetime $\tau$ of $T_{bb;\bar{u}
\bar{d} }^{-}$ are evaluated using its semileptonic decay channels $T_{bb;%
\bar{u} \bar{d}}^{-} \to Z_{bc;\bar{u} \bar{d}}^{0}l \bar{\nu_l}$, $l=e,\mu$
and $\tau$. The obtained results, $\Gamma=(7.17\pm 1.23)\times 10^{-8\ \ }%
\mathrm{MeV}$ and $\tau =9.18_{-1.34}^{+1.90}~\mathrm{fs}$, can be useful
for experimental investigations of the doubly-heavy tetraquarks.
\end{abstract}

\maketitle


\section{Introduction}

\label{sec:Int} 
Assumptions about the existence of four-quark bound states (tetraquarks)
were made in an early stage of QCD and aimed to explain some of the unusual
features of meson spectroscopy. Thus, the nonet of light scalar mesons
was considered as bound states of four light quarks rather than being
composed of a quark and an antiquark, as in the standard models of the
mesons. The stability problems of heavy and heavy-light tetraquarks were
also among the questions addressed in these studies \cite%
{Jaffe:1976ig,Jaffe:1976ih,Weinstein:1982gc,Ader:1981db}.

Due to the impressive experimental discoveries and theoretical progress of the past
$15$ years the study of multiquark hadrons has become an integral part of
high energy physics. During this period of development and growth  various
difficulties in experimental studies, and the classification and theoretical
interpretation of numerous tetraquarks were successfully overcome \cite%
{Chen:2016qju,Esposito:2016noz,Ali:2017jda,Olsen:2017bmm}.

But there are still problems in the physics of  exotic hadrons that are
not fully solved; the identification of the tetraquark resonances and their
stability are among these questions. It is known that the first
charmonium-like resonances observed experimentally were interpreted not only
as tetraquarks, but also as excited states of the conventional
charmonium. Fortunately, there are different classes of tetraquarks that
cannot be identified as charmonia or bottomonia states. Indeed, charged
resonances carrying one or two units of electric charge and states containing
two or more open quark flavors can easily be distinguished from charmonium-
or bottomonium-like structures. All of the resonances observed in various
experiments and classified as tetraquarks are unstable with respect to
strong interactions. They lie either above the open charm (-bottom)
thresholds or are very close to them. Such  four-quark compounds can strongly decay
 to two conventional mesons. Because quarks
required to create these mesons already exist  in the master particles,
the width of such states is rather large: the dissociation into two mesons
is the main strong decay channel of the unstable tetraquarks.

It is natural that theoretical explorations of stable four-quark systems and
their experimental discovery remain on the agenda of particle physics.
The tetraquarks built of heavy $cc$ or $bb$ diquarks and light antidiquarks
are real candidates for such states. Their studies have a long history; in
fact, the class of exotic mesons $QQ\bar{Q}\bar{Q}$ and $QQ\bar{q}\bar{q}$
were studied in Refs.\ \cite{Ader:1981db,Lipkin:1986dw,Zouzou:1986qh}, where
a potential model with an additive pairwise interaction was used to search
for stable tetraquarks. It was demonstrated that in the context of this
approach the exotic mesons composed of only heavy quarks are unstable, but
the tetraquarks $QQ\bar{q}\bar{q}$ may form  stable compounds provided
the ratio $m_{Q}/m_{q}$ is large. The same conclusions were made in Ref.\
\cite{Carlson:1987hh}, in which the only constraint imposed on the confining
potential was its finiteness when two particles come close together. There it was found that
the isoscalar $J^{P}=1^{+}$
tetraquark $T_{bb;\bar{u}\bar{d}}^{-}$ resides below the two-B-meson
threshold, and hence can decay only weakly. At the same time, the
tetraquarks $T_{cc;\bar{q}\bar{q}^{\prime }}$ and $T_{bc;\bar{q}\bar{q}%
^{\prime }}$ may exist as unstable or stable bound states. The stability of
the $QQ\overline{q}\overline{q}$ compounds in the limit $m_{Q}\rightarrow
\infty $ was studied in Ref.\ \cite{Manohar:1992nd}, as well.

Various theoretical models-starting from the chiral and dynamical quark
models and ending with the relativistic quark model-were used to study
the properties and compute the masses of the $T_{QQ}$ states \cite%
{Pepin:1996id,Janc:2004qn,Cui:2006mp,Vijande:2006jf,Ebert:2007rn}. The
masses of the axial-vector states $T_{QQ;\bar{u}\bar{d}}$ were also extracted
from the two-point sum rules \cite{Navarra:2007yw}. In accordance with
the results of Ref.\ \cite{Navarra:2007yw}, the mass of the tetraquark $T_{bb;%
\bar{u}\bar{d}}^{-}$ is $10.2\pm 0.3\ \mathrm{GeV}$, which is below
the open-bottom threshold. Using the same method, the parameters of
the $QQ\bar{q}\bar{q}$ states with the spin-parities $0^{-},\ 0^{+},\ 1^{-}$
and $1^{+}$ were evaluated in Ref.\ \cite{Du:2012wp}. The production
mechanisms of the $T_{cc}$ tetraquarks-such as the heavy ion and
proton-proton collisions, electron-positron annihilations, $B_{c}$ meson and
heavy $\Xi _{bc\text{ }}$ baryon decays-as well as possible decay channels
of the $T_{cc}$ states were addressed in the literature \cite%
{SchaffnerBielich:1998ci,DelFabbro:2004ta,Lee:2007tn,Hyodo:2012pm,Esposito:2013fma}.

The discovery of the doubly charmed baryon $\Xi _{cc}^{++}=ccu$ by the LHCb
Collaboration \cite{Aaij:2017ueg} inspired new investigations of
double-charm, double-bottom and four-bottom tetraquarks \cite%
{Karliner:2017qjm,Luo:2017eub,Eichten:2017ffp,Wang:2017dtg,Ali:2018ifm,
Ali:2018xfq,Eichten:2017ual,Hughes:2017xie,Esposito:2018cwh}. Lattice
simulations in the context of  nonrelativistic QCD to search for the
existence of the bound states $T_{bb;\overline{b}\overline{b}}^{0}$ below
the lowest bottomonium-pair threshold were carried out in Ref. \cite%
{Hughes:2017xie}, but no evidence was found for such stable states
with quantum numbers $0^{++},\ 1^{+-}$ and $2^{++}$, which can be considered
a present-day confirmation of the conclusions originally made in Refs. \cite%
{Ader:1981db,Lipkin:1986dw,Zouzou:1986qh,Carlson:1987hh}. A situation with
double-bottom tetraquarks is more promising. Thus, the mass of the state $%
T_{bb;\overline{u}\overline{d}}^{-}$ was estimated once more in the
framework of a phenomenological model in Ref.\ \cite{Karliner:2017qjm}. There,
 the mass of the isoscalar axial-vector
state $T_{bb;\overline{u}\overline{d}}^{-}$ was found to be  $m=10389\pm 12\
\mathrm{MeV}$ which is $215\ \mathrm{MeV}$ below the $B^{-}\overline{B}%
^{\ast 0}$ threshold and $170\ \mathrm{MeV}$ below the threshold for  $%
B^{-}\overline{B}^{0}\gamma $ decay. This means that the tetraquark $T_{bb;%
\overline{u}\overline{d}}^{-}$ is stable against the strong and
electromagnetic decays and only decays weakly. At the same time, the mass
of the double-charm $T_{cc;\overline{u}\overline{d}}^{+}$ state is $%
3882\pm 12\ \mathrm{MeV}$, which is above the thresholds of both  $%
D^{0}D^{\ast +}$ and $D^{0}D^{+}\gamma $ decays (see Ref. \ \cite{Karliner:2017qjm}).
The double-charm states $T_{cc;\overline{s}\overline{s}%
}^{++}$ and $T_{cc;\overline{d}\overline{s}}^{++}$ that belong to the class
of doubly charged tetraquarks were investigated recently in our work \cite%
{Agaev:2018vag}. These particles carry two units of electric charge, which makes them particularly interesting.
They are above the $D_{s}^{+}D_{s0}^{\ast +}(2317)$ and $D^{+}D_{s0}^{\ast +}(2317)$ thresholds,
and the width of the strong decays $T_{cc;\overline{s}\overline{s}%
}^{++}\rightarrow D_{s}^{+}D_{s0}^{\ast +}(2317)$ and $T_{cc;\overline{d}%
\overline{s}}^{++}\rightarrow D^{+}D_{s0}^{\ast +}(2317)$ allowed us to
classify them as relatively broad resonances.

In  light of recent progress made in the physics of double-heavy
tetraquarks and the expected stability of the $T_{bb;\overline{u}\overline{d}%
}^{-}$ state, its weak decays are a  very interesting subject for a detailed
analysis. The semileptonic decays of four-quark systems- when an initial
tetraquark transforms into a final tetraquark and $l\overline{\nu }_{l}$ or $%
\overline{l}\nu _{l}$ leptons- are a relatively new topic in the physics of
exotic mesons \cite{Sundu:2018uyi,Xing:2018bqt}. In Ref.\ \cite{Sundu:2018uyi}
the decay of the axial-vector tetraquark $Z_{s}=[cs][\overline{b}\overline{s}]$ to a final
state $X(4274)\overline{l}\nu _{l}$ was studied using the QCD sum rule method.
The widths of these decays, (where $l=e,\mu $
and $\tau $) are very small, and therefore the transitions $Z_{s}\rightarrow X(4274)%
\overline{l}\nu _{l}\ $ were classified  as
rare processes. The semileptonic decays of the stable double heavy
tetraquarks were considered in Ref.\ \cite{Xing:2018bqt}.

In the present work we are going to explore the semileptonic decays of the
tetraquark $T_{bb;\overline{u}\overline{d}}^{-}$ and evaluate its full width
and mean lifetime. The tetraquark $T_{bb;\overline{u}\overline{d}}^{-}$
undergoes  weak decay through the transition $b\rightarrow W^{-}c$. In
the final state, its decay products consist of $l\overline{\nu }_{l}$ and
a diquark-antidiquark $Z_{bc;\bar{u}\bar{d}}^{0}=[bc][\overline{u}\overline{d}%
] $ state (for simplicity, hereafter $Z_{bc}^{0}$). The tetraquark $%
Z_{bc}^{0}$ may decay to $B$ and $D$ mesons with appropriate masses and
spin-parities provided its mass is larger than corresponding thresholds. In
this scenario $Z_{bc}^{0}$ dissociates strongly to the final conventional
mesons. Otherwise, at the next stage $Z_{bc}^{0}$ should decay due to weak
or electromagnetic interactions. In the present work we restrict ourselves
by considering the semileptonic decay of $T_{bb;\overline{u}\overline{d}%
}^{-} $ only to the scalar state $Z_{bc}^{0}$.

The open charm-bottom four-quark systems $QQ^{\prime }\bar{q}\bar{q}$ were
already analyzed in Refs.\ \cite{Zouzou:1986qh,SilvestreBrac:1993ry}. In
recent investigations these compounds were treated either as $B_{c}$-like
molecular or $Z_{bc}=[bc][\overline{q}\overline{q}]$-type
diquark-antidiquark states. The masses of the $B_{c}$-like scalar and
axial-vector molecules with different light-quark contents and spin-parities
were calculated in Refs.\cite{Sun:2012sy,Albuquerque:2012rq}. The open charm-bottom states were
analyzed in Ref.\ \cite{Chen:2013aba} in the
framework of the diquark-antidiquark model. In order to extract the masses of these
states, the authors utilized the QCD sum rule method and interpolating currents
of different color structure. The class of open charm-bottom tetraquarks
also includes states with $(b,\overline{c})$ or $(c,\overline{b})$ quarks
which were the subject of rather intensive studies as well \cite%
{Sun:2012sy,Albuquerque:2012rq,Chen:2013aba,Zhang:2009vs,Zhang:2009em,Agaev:2016dsg,Agaev:2017uky}%
. In fact, the molecule-type tetraquarks with the contents $\{Q\overline{q}%
\}\{\overline{Q}^{(\prime )}q\}$ and $\{Q\overline{s}\}\{\overline{Q}%
^{(\prime )}s\}$ were studied in Refs.\ \cite{Zhang:2009vs} and \cite%
{Zhang:2009em}, respectively. In these papers the masses of these
hypothetical particles were computed in the context of the QCD two-point sum
rule approach using vacuum condensates up to dimension six. The
spectroscopic parameters and strong decays of the scalar and axial-vector
tetraquarks $Z_{q}=[cq][\overline{b}\overline{q}]$ and $Z_{s}=[cs][\overline{%
b}\overline{s}]$ were calculated in Refs.\ \cite{Agaev:2016dsg} and \cite%
{Agaev:2017uky}, respectively.

It is remarkable that $Z_{bc}^{0}=[bc][\overline{u}\overline{d}]$ is the
open charm-bottom tetraquark, and that it contains four quarks of
different flavors. Two years ago, data on the state known as $X(5568)$ from the
D0 Collaboration \cite{D0:2016mwd}
led to an interest in compound systems of four distinct quarks. However,
both the experimental and theoretical studies of  $X(5568)$
led to controversial conclusions, leaving the status of this tetraquark
unclear. Therefore investigating the process $T_{bb;\overline{u}\overline{%
d}}^{-}\rightarrow Z_{bc}^{0}l\overline{\nu }_{l}$ could not only  help  to answer
questions about features of the tetraquark $T_{bb;\overline{u}%
\overline{d}}^{-}$ itself, but also to clarify the structure and properties of
its decay products.

The spectroscopic parameters of $T_{bb;\overline{u}\overline{d}}^{-}$ and $%
Z_{bc}^{0}$ are important input for studying  the semileptonic
decay under consideration. In the present work, we calculate the masses and
couplings of these tetraquarks by employing QCD sum rules obtained from
an analysis of the relevant two-point correlation functions. When computing
the correlation functions, we take into account the vacuum expectation values of
the quark, gluon, and mixed local operators up to dimension ten. We
evaluate the width of the semileptonic decay $T_{bb;\overline{u}\overline{d}%
}^{-}\rightarrow Z_{bc}^{0}l\overline{\nu }_{l}$ by applying the standard
prescriptions of the QCD three-point sum rule method. Our aim here is to
extract the sum rules for the weak form factors $G_{i}(q^{2}),\,i=0,1,2,3$
and to compute their numerical values. This allows us to determine the so-called
fit functions $F_{i}(q^{2})$, which coincide with $G_{i}(q^{2})$, but can be
extended to a  region of momentum transfers that is not accessible to
the QCD sum rules. The functions $F_{i}(q^{2})$ are used to integrate the
differential decay rate $d\Gamma /dq^{2}$ and find the partial width of the
decay processes $\Gamma \left( T_{bb;\overline{u}\overline{d}%
}^{-}\rightarrow Z_{bc}^{0}l\overline{\nu }_{l}\right) $, $l=e,\ \mu $ and $%
\tau $.

This article is organized in the following manner: In Sec.\ \ref{sec:Masses}
we derive the QCD two-point sum rules for the masses and couplings of the
tetraquarks $T_{bb;\overline{u}\overline{d}}^{-}$ and $Z_{bc}^{0}$, and
numerically compute their values. In Sec.III we use
the QCD three-point correlation function  to derive sum rules for
the weak form factors $G_{i}(q^{2})$. In this section we also perform a
numerical analysis of the obtained sum rules and determine the fit functions,
which allow us to evaluate the width of the semileptonic decay $T_{bb;%
\overline{u}\overline{d}}^{-}\rightarrow Z_{bc}^{0}l\overline{\nu }_{l}$ and
mean lifetime of the state $T_{bb;\overline{u}\overline{d}}^{-}$. Section IV
 contains a discussion of the obtained results and our brief
conclusions. The explicit expression for the decay rate $d\Gamma /dq^{2}$ can be found
in the Appendix.


\section{Spectroscopic parameters of the tetraquarks $T_{bb;\overline{u}%
\overline{d}}^{-}$ and $Z_{bc}^{0}$}

\label{sec:Masses}

In this section we calculate the spectroscopic parameters of the tetraquarks
$T_{bb;\overline{u}\overline{d}}^{-}$ and $Z_{bc}^{0}$ by employing the QCD
two-point sum rules extracted from analysis of the relevant correlation
functions $\Pi _{\mu \nu }(p)$ and $\Pi (p)$. The masses of $T_{bb;\overline{u}\overline{d}}^{-}$
and $Z_{bc}$ in the framework of QCD sum rules  were found
in Refs. \cite{Navarra:2007yw,Du:2012wp} and \cite%
{Chen:2013aba}, respectively. We are going to evaluate the masses and
tetraquark-current couplings of these states by taking into account the
vacuum condensates up to dimension ten which exceeds the accuracy of the
previous studies:  updated information on the spectroscopic parameters of
the tetraquarks $T_{bb;\overline{u}\overline{d}}^{-}$ and $Z_{bc}^{0}$ is
necessary to explore the semileptonic decay $T_{bb;\overline{u}\overline{d}%
}^{-}\rightarrow Z_{bc}^{0}l\overline{\nu }_{l}$ in the next section.

The function $\Pi _{\mu \nu }(p)$ is defined as
\begin{equation}
\Pi _{\mu \nu }(p)=i\int d^{4}xe^{ip\cdot x}\langle 0|\mathcal{T}\{J_{\mu
}(x)J_{\nu }^{\dag }(0)\}|0\rangle ,  \label{eq:CF1}
\end{equation}%
where $J_{\mu }(x)$ is the interpolating current to the axial-vector
tetraquark $T_{bb;\overline{u}\overline{d}}^{-}$ composed of an axial-vector
diquark and a scalar antidiquark. This current is given by  \cite%
{Navarra:2007yw}
\begin{equation}
J_{\mu }(x)=b_{a}^{T}(x)C\gamma _{\mu }b_{b}(x)\overline{u}_{a}(x)\gamma
_{5}C\overline{d}_{b}^{T}(x).  \label{eq:Curr1}
\end{equation}%
Here, $a$ and $b$ are the color indices and $C$ is the charge-conjugation
operator.

The correlation function $\Pi (p)$ for the scalar tetraquark $Z_{bc}^{0}$
has the form
\begin{equation}
\Pi (p)=i\int d^{4}xe^{ip\cdot x}\langle 0|\mathcal{T}\{J^{Z}(x)J^{Z\dag
}(0)\}|0\rangle ,  \label{eq:CF2}
\end{equation}%
where the current $J^{Z}(x)$ is defined as
\begin{eqnarray}
J^{Z}(x) &=&b_{a}^{T}(x)C\gamma _{5}c_{b}(x)\left[ \overline{u}_{a}(x)\gamma
_{5}C\overline{d}_{b}^{T}(x)\right.   \notag \\
&&\left. -\overline{u}_{b}(x)\gamma _{5}C\overline{d}_{a}^{T}(x)\right],
\label{eq:Curr2}
\end{eqnarray}%
and is obtained using currents for the diquark-antidiquarks $Z_{bc}$ from Ref.
\cite{Chen:2013aba}. The current $J^{Z}(x)$ is composed of a scalar diquark and an antidiquark
in the antitriplet and triplet representations of the color group, respectively.

Here we concentrate  on calculating  the parameters of the tetraquark $%
T_{bb;\overline{u}\overline{d}}^{-}$ and only provide  necessary expressions
and final results for $Z_{bc}^{0}$. In accordance with QCD sum rule method
one first has to express the correlation function $\Pi _{\mu \nu }(p)$ in
terms of the tetraquarks' mass $m$ and coupling $f$, which form the
phenomenological or physical side of the sum rules. We treat the tetraquark $%
T_{bb;\overline{u}\overline{d}}^{-}$ as a ground-state particle in its
class, and therefore we isolate only the first term in $\Pi _{\mu \nu }^{\mathrm{Phys}%
}(p)$ which is given by
\begin{equation}
\Pi _{\mu \nu }^{\mathrm{Phys}}(p)=\frac{\langle 0|J_{\mu }|T(p)\rangle
\langle T(p)|J_{\nu }^{\dagger }|0\rangle }{m^{2}-p^{2}}+\dots .
\label{eq:CF3}
\end{equation}%
This expression is derived by saturating the correlation function  (\ref%
{eq:CF1}) with a complete set of states with $J^{P}=1^{+}$ and performing
the integration over $x$. The dots here indicate contributions to $\Pi _{\mu
\nu }^{\mathrm{Phys}}(p)$ from higher resonances and continuum states.

The function $\Pi _{\mu \nu }^{\mathrm{Phys}}(p)$ can be further simplified
by introducing the matrix element
\begin{equation}
\langle 0|J_{\mu }|T(p,\epsilon )\rangle =fm\epsilon _{\mu },
\label{eq:MElem1}
\end{equation}%
where $\epsilon _{\mu }$ is the polarization vector of the $T_{bb;\overline{u%
}\overline{d}}^{-}$ state. It is not difficult to demonstrate that in terms
of $m$ and $f$ the function takes the following form
\begin{equation}
\Pi _{\mu \nu }^{\mathrm{Phys}}(p)=\frac{m^{2}f^{2}}{m^{2}-p^{2}}\left(
-g_{\mu \nu }+\frac{p_{\mu }p_{\nu }}{m^{2}}\right) +\ldots
\label{eq:CorM}
\end{equation}%
To suppress the contribution arising from the higher resonances and
continuum, we carry out the Borel transformation of the correlation
function, which reads
\begin{equation}
\mathcal{B}\Pi _{\mu \nu }^{\mathrm{Phys}}(p)=m^{2}f^{2}e^{-m^{2}/M^{2}}%
\left( -g_{\mu \nu }+\frac{p_{\mu }p_{\nu }}{m^{2}}\right) +\ldots ,
\label{eq:CF4}
\end{equation}%
where $M^{2}$ is the Borel parameter.

The second part of the sum rules is given by the same correlation function $%
\Pi _{\mu \nu }(p)$, but expressed in terms of the quark propagators
\begin{eqnarray}
&&\Pi _{\mu \nu }^{\mathrm{OPE}}(p)=i\int d^{4}xe^{ip\cdot x}\left\{ \mathrm{%
Tr}\left[ \gamma _{5}\widetilde{S}_{d}^{b^{\prime }b}(-x)\gamma
_{5}S_{u}^{a^{\prime }a}(-x)\right] \right.  \notag \\
&&\times \mathrm{Tr}\left[ \gamma _{\nu }\widetilde{S}_{b}^{aa^{\prime
}}(x)\gamma _{\mu }S_{b}^{bb^{\prime }}(x)\right] -\mathrm{Tr}\left[ \gamma
_{5}\widetilde{S}_{d}^{b^{\prime }b}(-x)\right.  \notag \\
&&\left. \left. \times \gamma _{5}S_{u}^{a^{\prime }a}(-x)\right] \mathrm{Tr}%
\left[ \gamma _{\nu }\widetilde{S}_{b}^{ba^{\prime }}(x)\gamma _{\mu
}S_{b}^{ab^{\prime }}(x)\right] \right\} .  \label{eq:CF5}
\end{eqnarray}%
In Eq.\ (\ref{eq:CF5}) $S_{b}^{ab}(x)$ and $S_{q}^{ab}(x)$ are the $b$ and $%
q(u,d)$-quark propagators, explicit expression for which can be found, for
example, in Ref.\ \cite{Sundu:2018uyi}. Here we also introduce the notation
\begin{equation}
\widetilde{S}_{b(q)}(x)=CS_{b(q)}^{T}(x)C.  \label{eq:Prop}
\end{equation}

The QCD sum rules can be extracted by using the same Lorentz structures in
both  $\Pi _{\mu \nu }^{\mathrm{Phys}}(p)$ and $\Pi _{\mu \nu }^{\mathrm{%
OPE}}(p)$. The structures $\sim g_{\mu \nu }$ are appropriate for our
purposes, because they receive contributions only from spin-$1$ particles.
The invariant amplitude $\Pi ^{\mathrm{OPE}}(p^{2})$ corresponding to this
structure can be represented by the dispersion integral
\begin{equation}
\Pi ^{\mathrm{OPE}}(p^{2})=\int_{4m_{b}^{2}}^{\infty }\frac{\rho ^{\mathrm{%
OPE}}(s)}{s-p^{2}}ds+\ldots ,  \label{eq:Ampl1}
\end{equation}%
where $\rho ^{\mathrm{OPE}}(s)$ is the two-point spectral density. It is
proportional to the imaginary part of the structure $\sim g_{\mu \nu }$ in
the function $\Pi _{\mu \nu }^{\mathrm{OPE}}(p).$ In the present work, $\rho
^{\mathrm{OPE}}(s)$ is calculated by taking into account the quark, gluon,
and mixed vacuum condensates up to dimension ten.

By applying the Borel transformation to $\Pi ^{\mathrm{OPE}}(p^{2})$,
equating the obtained expression with the relevant part of the function $%
\mathcal{B}\Pi _{\mu \nu }^{\mathrm{Phys}}(p)$, and performing the continuum
subtraction we find the final sum rules. Then, the mass of the $T_{bb%
\overline{u}\overline{d}}^{-}$ state can be evaluated from the sum rule
\begin{equation}
m^{2}=\frac{\int_{4m_{b}^{2}}^{s_{0}}dss\rho ^{\mathrm{OPE}}(s)e^{-s/M^{2}}}{%
\int_{4m_{b}^{2}}^{s_{0}}ds\rho ^{\mathrm{OPE}}(s)e^{-s/M^{2}}},
\label{eq:Mass1}
\end{equation}%
whereas to find the coupling $f$ we employ the expression
\begin{equation}
f^{2}=\frac{1}{m^{2}}\int_{4m_{b}^{2}}^{s_{0}}ds\rho ^{\mathrm{OPE}%
}(s)e^{(m^{2}-s)/M^{2}}.  \label{eq:Coupl1}
\end{equation}%
Here $s_{0}$ is the continuum threshold parameter that separates the
ground-state and continuum contributions from one another.

In the case of the scalar tetraquark $Z_{bc}^{0}$, there are some differences
stemming from its spin-parity and the structure of the interpolating
current. Thus, the matrix element $\langle 0|J^{Z}|Z(p)\rangle $ has the
form
\begin{equation}
\langle 0|J^{Z}|Z(p)\rangle =f_{Z}m_{Z},  \label{eq:MElem2}
\end{equation}%
which is analogous to the matrix element of a conventional scalar meson. The
correlation function $\Pi ^{\mathrm{OPE}}(p)$ is given by
\begin{eqnarray}
&&\Pi ^{\mathrm{OPE}}(p)=i\int d^{4}xe^{ip\cdot x}\mathrm{Tr}\left[
S_{c}^{bb^{\prime }}(x)\gamma _{5}\widetilde{S}_{b}^{aa^{\prime }}(x)\gamma
_{5}\right]  \notag \\
&&\times \left\{ \mathrm{Tr}\left[ \gamma _{5}\widetilde{S}_{d}^{b^{\prime
}b}(-x)\gamma _{5}S_{u}^{a^{\prime }a}(-x)\right] -\mathrm{Tr}\left[ \gamma
_{5}\widetilde{S}_{d}^{a^{\prime }b}(-x)\right. \right.  \notag \\
&&\left. \times \gamma _{5}S_{u}^{b^{\prime }a}(-x)\right] -\mathrm{Tr}\left[
\gamma _{5}\widetilde{S}_{d}^{b^{\prime }a}(-x)\gamma _{5}S_{u}^{a^{\prime
}b}(-x)\right]  \notag \\
&&\left. +\mathrm{Tr}\left[ \gamma _{5}\widetilde{S}_{d}^{a^{\prime
}a}(-x)\gamma _{5}S_{u}^{b^{\prime }b}(-x)\right] \right\} .  \label{eq:CF6}
\end{eqnarray}%
The remaining manipulations and final sum rules for $m_{Z}$ and $f_{Z}$ are
similar to those for the tetraquark $T_{bb;\overline{u}\overline{d}}^{-}$.

\begin{table}[tbp]
\begin{tabular}{|c|c|}
\hline\hline
Parameters & Values \\ \hline\hline
$m_{b}$ & $4.18^{+0.04}_{-0.03}~\mathrm{GeV}$ \\
$m_{c}$ & $(1.27 \pm 0.03)~\mathrm{GeV}$ \\
$\langle \bar{q}q \rangle $ & $-(0.24\pm 0.01)^3$ $\mathrm{GeV}^3$ \\
$\langle \bar{s}s \rangle $ & $0.8\ \langle \bar{q}q \rangle$ \\
$m_{0}^2 $ & $(0.8\pm0.1)$ $\mathrm{GeV}^2$ \\
$\langle \overline{q}g_{s}\sigma Gq\rangle$ & $m_{0}^2\langle \bar{q}q
\rangle $ \\
$\langle \overline{s}g_{s}\sigma Gs\rangle$ & $m_{0}^2\langle \bar{s}s
\rangle $ \\
$\langle\frac{\alpha_sG^2}{\pi}\rangle $ & $(0.012\pm0.004)$ $~\mathrm{GeV}%
^4 $ \\
$\langle g_{s}^3G^3\rangle $ & $(0.57\pm0.29)$ $~\mathrm{GeV}^6 $ \\
\hline\hline
\end{tabular}%
\caption{The parameters utilized in numerical computations.}
\label{tab:Param}
\end{table}
The obtained sum rules depend on the quark, gluon, and mixed condensates,
the numerical values of which are collected in Table\ \ref{tab:Param}. This
table  also contains the masses of the $b$ and $c$ quarks, which appear in the
sum rules as input parameters.

Besides, Eqs.\ (\ref{eq:Mass1}) and (\ref{eq:Coupl1}) depend on the auxiliary
parameters $M^{2}$ and $s_{0}$, which should satisfy the standard constraints of
the sum rule computations. Our analysis proves that the working windows
\begin{equation}
M^{2}\in \lbrack 9,~13]\ \mathrm{GeV}^{2},\ s_{0}\in \lbrack 115,~120]\
\mathrm{GeV}^{2}  \label{eq:Reg1}
\end{equation}%
meet all of the restrictions imposed on $M^{2}$ and $s_{0}$. Thus, the maximum of
the Borel parameter is determined from the minimum allowed value of the pole
contribution ($\mathrm{PC}$), which at $M^{2}=13~\mathrm{GeV}^{2}$ is $%
16\%$ of the full correlation function. Within the region $M^{2} \in \lbrack
9,~13]~\mathrm{GeV}^{2}$ the pole contribution varies from $59$ to $16\%$.
The lower limit of the Borel parameter is fixed by the convergence of the
operator product expansion (OPE) for the correlation function. In the
present work, we use the criterion%
\begin{equation}
R(M^{2})=\frac{\Pi ^{\mathrm{Dim(8+9+10)}}(M^{2},\ s_{0})}{\Pi (M^{2},\
s_{0})}<0.05,  \label{eq:Conv}
\end{equation}%
where $\Pi (M^{2},\ s_{0})$ is the Borel-transformed and subtracted function
$\Pi ^{\mathrm{OPE}}(p^{2})$, and $\Pi ^{\mathrm{Dim(8+9+10)}}(M^{2},\
s_{0}) $ is  the contribution from the last three terms in its expansion. At $%
M^{2}=9~\mathrm{GeV}^{2}$ the ratio $R$ is equal to $R(9~\mathrm{GeV}%
^{2})=0.01$, which ensures the excellent convergence of the sum rules.
Moreover, at $M^{2}=9~\mathrm{GeV}^{2}$ the perturbative contribution
amounts to $74\%$ of the full result,  considerably exceeding the
nonperturbative terms.

The quantities evaluated by means of the sum rules, in general, should not
depend on the auxiliary parameters $M^{2}$ and $s_{0}$. But in calculations
of the mass $m$ and coupling $f$ we observe a residual dependence on $M^{2}$
and $s_{0}$. Therefore, the stability of the extracted parameters (i.e., $m$ and $f$)
is a necessary condition to fix the working windows for $M^{2}$ and $s_{0}$.
In Figs.\ \ref{fig:Mass1} and \ref{fig:Coupl1} we plot the dependence of the
mass and coupling of the tetraquark $T_{bb;\overline{u}\overline{d}}^{-}$ on
the parameters $M^{2}$ and $s_{0}$. It is seen that $m$ and $f$ depend on $%
M^{2}$ and $s_{0}$, which generates the main part of the theoretical errors
inherent to the sum rule computations. For the mass $m$ these ambiguities
are small, whereas for the coupling $f$ they may be sizable. This behavior
has a simple explanation:  the sum rule for the mass of the tetraquark
(\ref{eq:Mass1}) is given as the ratio of integrals over the functions $%
s\rho ^{\mathrm{OPE}}(s)$ and $\rho ^{\mathrm{OPE}}(s)$, which considerably
reduces effects due to the variation of $M^{2}$ and $s_{0}$. The coupling $f$
depends on the integral over the spectral density $\rho ^{\mathrm{OPE}}(s)$
itself, and therefore undergoes relatively sizable changes. In the case
under discussion, theoretical errors for $m$ and $f$ stemming from
the uncertainties of $M^{2}$ and $s_{0}$ and other input parameters are
$ \pm 2.6$ and $ \pm 20\%$ of the corresponding central values,
respectively.

\begin{widetext}

\begin{figure}[h!]
\begin{center} \includegraphics[%
totalheight=6cm,width=8cm]{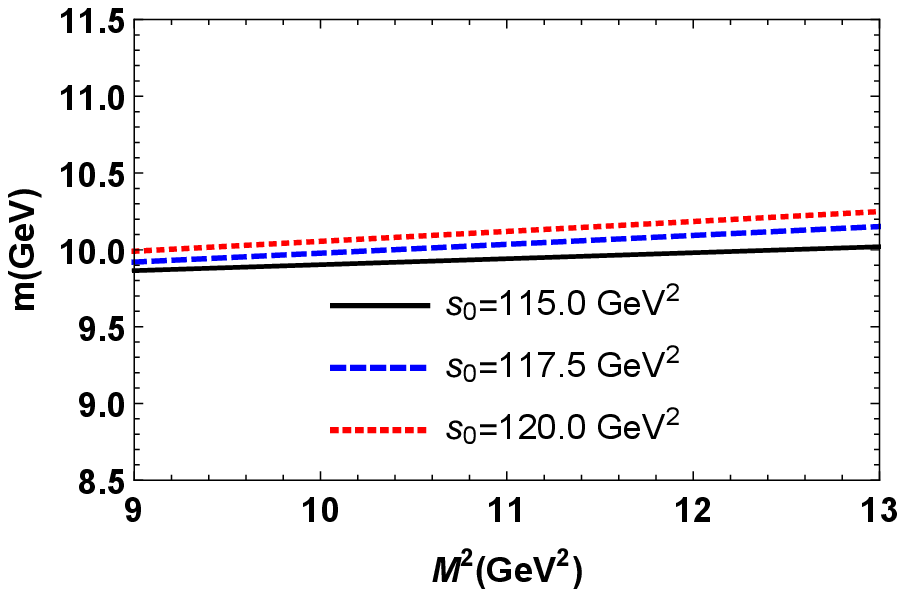}\,\,
\includegraphics[
totalheight=6cm,width=8cm]{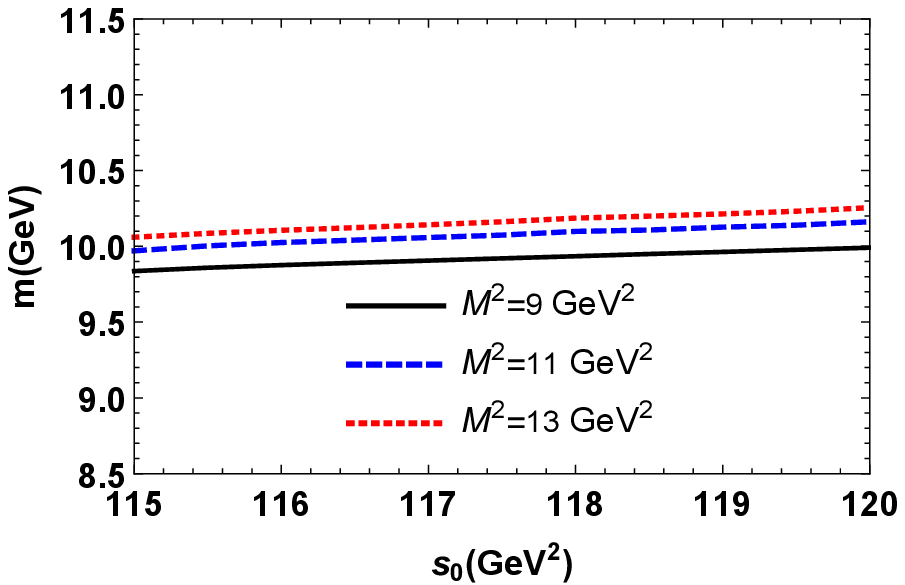}
\end{center}
\caption{ The mass of the tetraquark $T_{bb;\overline{u}%
\overline{d}}^{-}$ as a function of  the Borel parameter (left) and continuum threshold
parameter (right).}
\label{fig:Mass1}
\end{figure}

\begin{figure}[h!]
\begin{center} \includegraphics[%
totalheight=6cm,width=8cm]{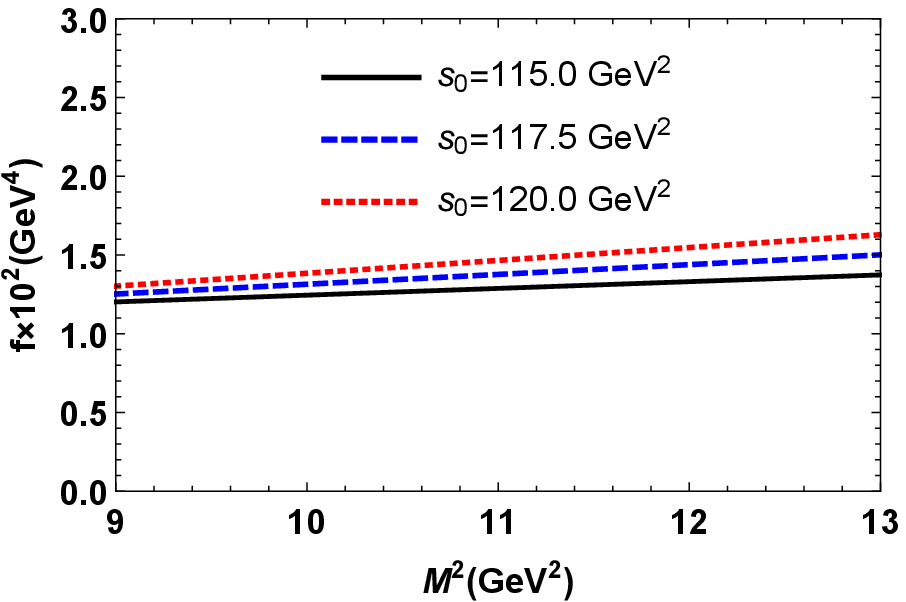}\,\,
\includegraphics[
totalheight=6cm,width=8cm]{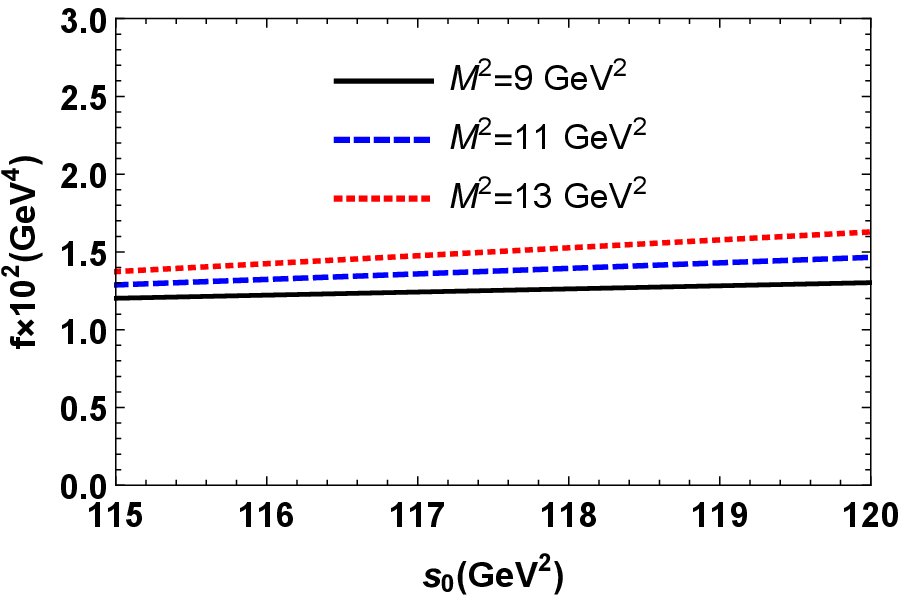}
\end{center}
\caption{ The coupling $f$ vs $M^2$ (left) and $s_0$ (right).}
\label{fig:Coupl1}
\end{figure}

\end{widetext}

Our analysis for the mass and coupling of the tetraquark $T_{bb;\overline{d}%
\overline{u}}^{-}$ predicts
\begin{eqnarray}
m &=&(10035~\pm 260)~\mathrm{MeV},  \notag \\
f &=&(1.38\pm 0.27)\cdot 10^{-2}\ \mathrm{GeV}^{4}.  \label{eq:CMass1}
\end{eqnarray}%
Similar studies of  $Z_{bc}^{0}$ lead to the following results:
\begin{eqnarray}
m_{Z} &=&(6660\pm 150~)\ \mathrm{MeV},  \notag \\
f_{Z} &=&(0.51\pm 0.16)\cdot 10^{-2}\ \mathrm{GeV}^{4},  \label{eq:CMass2}
\end{eqnarray}%
which have been obtained using the working regions
\begin{equation}
M^{2}\in \lbrack 5.5,\ 6.5]\ \mathrm{GeV}^{2},\ s_{0}\in \lbrack 53,\ 55]\
\mathrm{GeV}^{2}.  \label{eq:Reg2}
\end{equation}%
It is worth noting that in the calculations of $m_{Z}$ and $f_{Z}$ the
 $\mathrm{PC}$ by $55$ to $-21\%$. The contribution
of the last three terms to the corresponding correlation function at the
point $M^{2}=5.5\ \mathrm{GeV}^{2}$ amounts to $1.9\%$ of the total result,
which guarantees the convergence of the sum rules. In Figs.\ \ref{fig:Mass2}
and \ \ref{fig:Coupl2} we depict the mass and coupling of the tetraquark $%
Z_{bc}^{0}$ as a function of $M^{2}$ and $s_{0}$ to demonstrate their
residual dependence on these parameters. It is evident that, as in the case
of the $T_{bb;\overline{d}\overline{u}}^{-}$ state, the mass $m_{Z}$ is less
sensitive to variations of $M^{2}$ and $s_{0}$ than the coupling $f_{Z}$.
But, the relevant theoretical errors stay within the allowed limits inherent to sum rule computations,
which may equal up to $\pm 30\%$ of the predictions.

As it has been noted above, the mass of the state $T_{bb;\overline{u}%
\overline{d}}^{-}$ was evaluated using different approaches in Refs.\ \cite%
{Navarra:2007yw,Du:2012wp} and \cite{Karliner:2017qjm}. The investigations in
the first two papers were carried out in the framework of the sum rules
method, therefore we first compare our result for $m$ with those
predictions. Our result for $m$ is smaller than the prediction $m=10.2\pm 0.3\
\mathrm{GeV}$ made in Ref.\ \cite{Navarra:2007yw}: there is an overlapping
region between these two results, but the central values differ from each
other. This discrepancy is presumably connected  with the accuracy of the
analysis performed there (up to dimension-eight condensates), and with the
choice of the working intervals for the parameters $M^{2}$ and $s_{0}$.
Thus, in Ref. \cite{Navarra:2007yw} the explored range for the continuum
threshold was $11.3\leq \sqrt{s_{0}}\leq 11.7$ $\mathrm{GeV}$, whereas the
Borel parameter varied within the limits $M^{2}\in \lbrack 7.5,\ 9.6]\
\mathrm{GeV}^{2}$ or $M^{2}\in \lbrack 7.5,\ 11.2]\ \mathrm{GeV}^{2}$.
Because $\sqrt{s_{0}}$ determines the mass of the first excited tetraquark $%
T_{bb;\overline{u}\overline{d}}^{-}$ the corresponding mass gap amounts to $%
\Delta m=1.30\pm 0.36\ \mathrm{GeV}$ which is larger than the typical tetraquark
value $\Delta m_{T}\sim 0.5-0.7$ $\mathrm{GeV}$. In our case,
this mass gap is $\Delta m=0.79\pm 0.17\ \mathrm{GeV}$ and overshoots
$\Delta m_{T}$ as well. But one should take into account that the estimate $%
\Delta m_{T}\sim 0.6$ $\mathrm{GeV}$ was made for tetraquarks lying near or
above the corresponding two-meson thresholds, and therefore this fact may be
connected with the stable nature of $T_{bb;\overline{u}\overline{d}}^{-}$.

The sum rules analysis of the state $T_{bb;\overline{u}\overline{d}}^{-}$
was performed in Ref.\ \cite{Du:2012wp} by employing various interpolating
currents $\eta _{i}$. In computations the continuum threshold $s_{0}=115\
\mathrm{GeV}^{2}$ and different regions for the Borel parameter were used
with $M^{2}=[6.5,\ 8.6]\ \mathrm{GeV}^{2}$ and $M^{2}=[7.0,\ 9.2]\ \mathrm{%
GeV}^{2}$ being two extreme choices for $M^{2}$. The mass of the
axial-vector tetraquark $T_{bb;\overline{u}\overline{d}}^{-}$ in Ref.\ \cite{Du:2012wp}
was found to be $m=10.2\pm 0.3\ \mathrm{GeV}$. Here we also underline a
difference between the Borel windows in Ref. \cite{Du:2012wp} and those in the
present work as a possible source of this deviation.

The recent model analysis of Ref. \cite{Karliner:2017qjm} predicted $%
m=10389\pm 12\ \mathrm{MeV}$ which is considerably larger than the present
result. Nevertheless, all calculations confirm that the tetraquark $T_{bb;%
\overline{u}\overline{d}}^{-}$ is stable against the strong and
electromagnetic decays and can only dissociate  weakly.

The tetraquarks $Z_{bc}=[bc][\overline{q}\overline{q}]$ ($q=u,d$) were
investigated in Ref. \cite{Chen:2013aba} by employing the QCD sum rule
method and various interpolating currents. The masses of the charged scalar
tetraquarks $Z_{bc;\overline{u}\overline{u}}^{-}=[bc][\overline{u}\overline{%
u}]$ and $Z_{bc;\overline{d}\overline{d}}^{+}=[bc][\overline{d}\overline{d}]$
found there were $m=7.14\pm 0.10$ $\mathrm{GeV}$. This prediction is
considerably higher than our present result for $m_{Z}$. But one should take
into account that the scalar tetraquark $Z_{bc;\bar{u}\bar{d}}^{0}=[bc][%
\overline{u}\overline{d}]$ has different quark content: it is a neutral
particle and contains [like the resonance $X(5568)$]  four quarks of
different flavors. Therefore, a discrepancy between the  predictions for $%
Z_{bc}$ and $Z_{bc}^{0}$ may be explained not only by the accuracy of the
corresponding sum rule analysis and different working regions for the
parameters $M^{2}$ and $s_{0}$, but also by the aforementioned reasons. \
In Ref.\cite{Feng:2013kea} the masses of the ground-state tetraquarks $QQ^{\prime }\overline{u}%
\overline{d}$ in the context of the Bethe-Salpeter method.
In the case of the state $Z_{bc}^{0}$ using one of
parameters sets the authors found that its mass is  $m=6.93$ $%
\mathrm{GeV}$: this estimate is closer to our prediction.

\begin{widetext}

\begin{figure}[h!]
\begin{center} \includegraphics[%
totalheight=6cm,width=8cm]{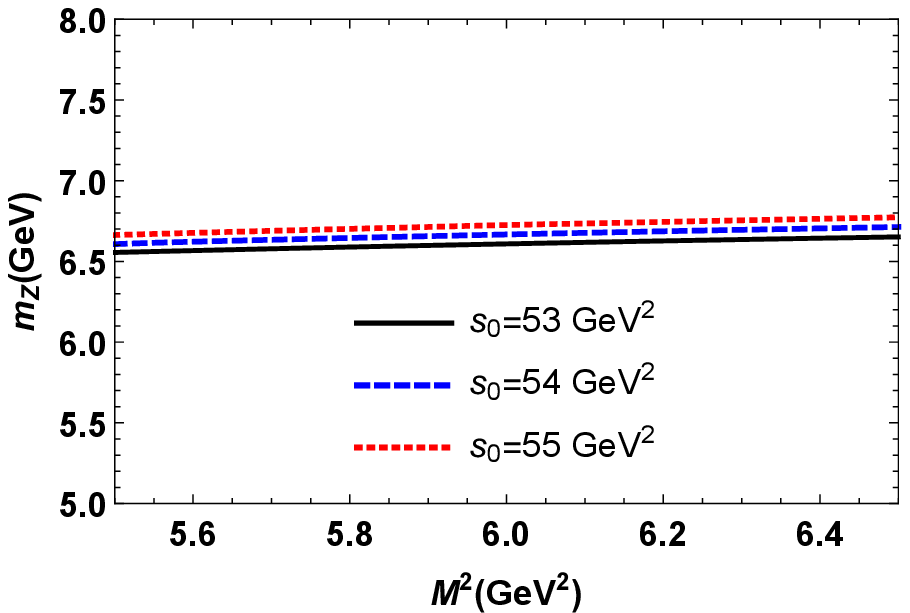}\,\,
\includegraphics[
totalheight=6cm,width=8cm]{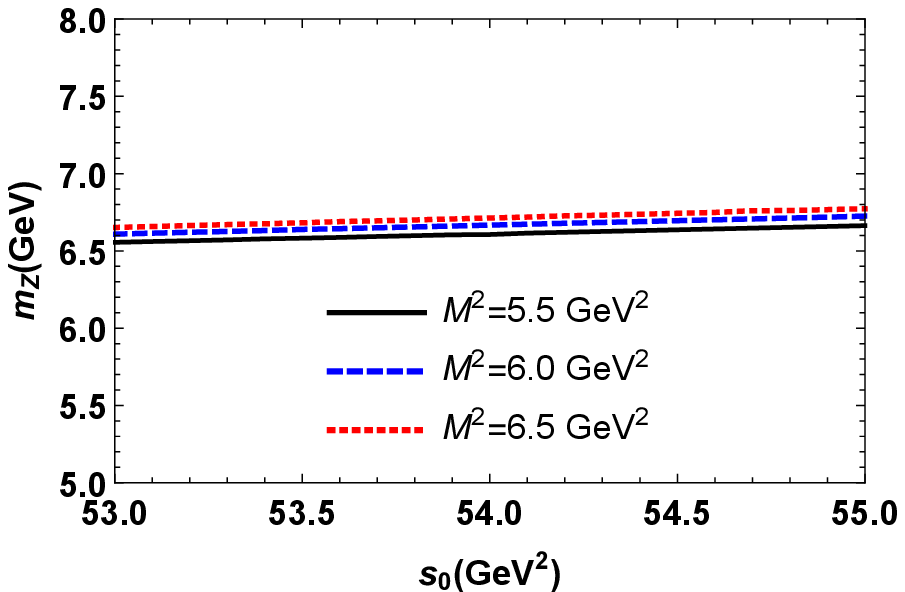}
\end{center}
\caption{ The same as in Fig. 1, but for the mass of the tetraquark $Z_{bc}^{0}$.}
\label{fig:Mass2}
\end{figure}

\begin{figure}[h!]
\begin{center} \includegraphics[%
totalheight=6cm,width=8cm]{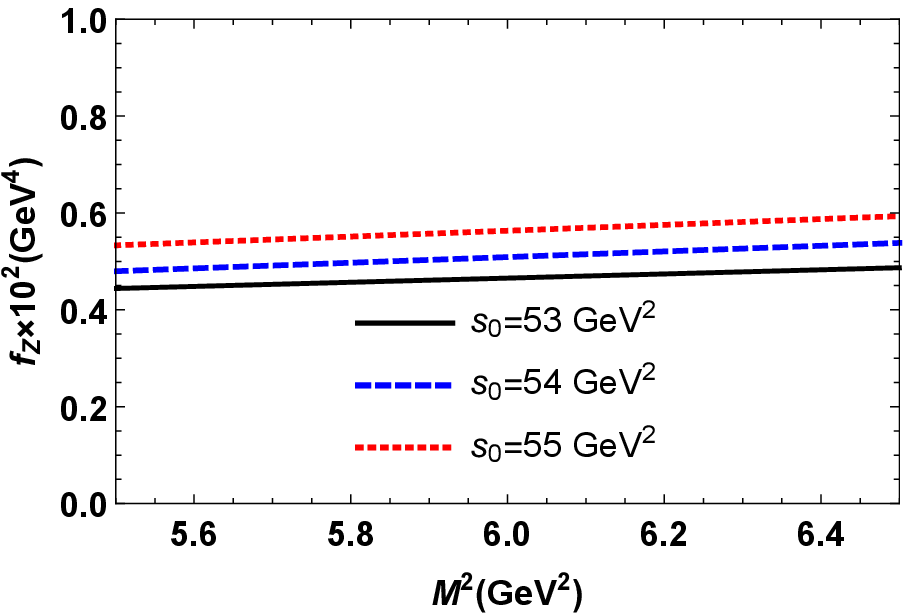}\,\,
\includegraphics[
totalheight=6cm,width=8cm]{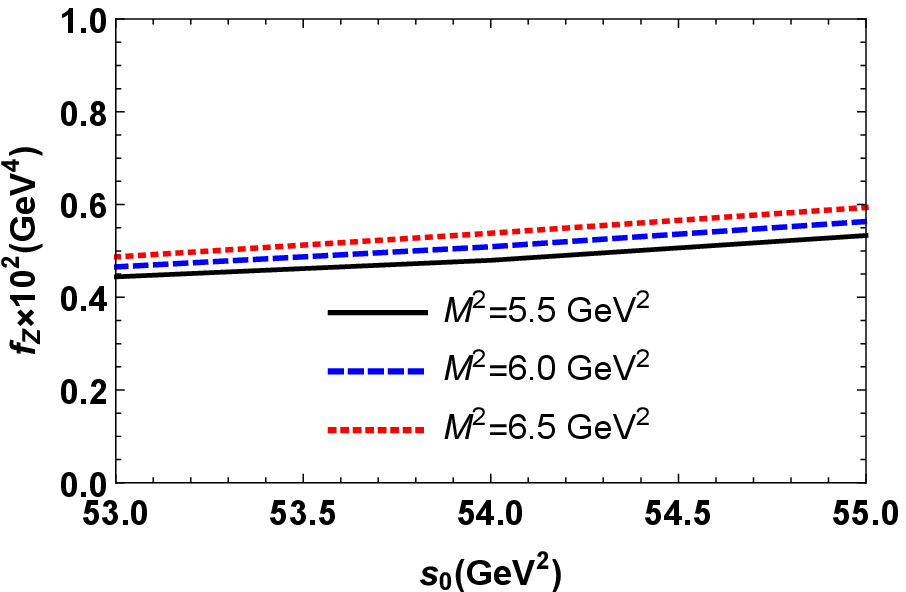}
\end{center}
\caption{ The coupling $f_Z$ of the tetraquark $Z_{bc}^{0}$ vs $M^2$ (left) and $s_0$ (right).}
\label{fig:Coupl2}
\end{figure}

\end{widetext}


\section{Semileptonic decay $T_{bb;\overline{u}\overline{d}}^{-}\rightarrow
Z_{bc}^{0}l\overline{\protect\nu }_{l}$}

\label{Decay}
The semileptonic decay of the tetraquark $T_{bb;\overline{u}\overline{d}%
}^{-} $ to the final state $Z_{bc}^{0}l\overline{\nu }_{l}$ runs through the
chain of transitions $b\rightarrow W^{-}c$ and $W^{-}\rightarrow l\overline{%
\nu }$. As is seen from results obtained in the previous section, the difference between the
initial and final tetraquarks masses is large enough to make all of the decays $%
l=e,\ \mu $ and $\tau $ kinematically allowed processes.

At the tree level the transition $b\rightarrow c$ can be described using the
effective Hamiltonian
\begin{equation}
\mathcal{H}^{\mathrm{eff}}=\frac{G_{F}}{\sqrt{2}}V_{bc}\overline{c}\gamma
_{\mu }(1-\gamma _{5})b\overline{l}\gamma ^{\mu }(1-\gamma _{5})\nu _{l},
\label{eq:EffH}
\end{equation}%
where $G_{F}$ is the Fermi coupling constant and $V_{bc}$ is the
corresponding element of the Cabibbo-Kobayashi-Maskawa (CKM) matrix. After
sandwiching the $\mathcal{H}^{\mathrm{eff}}$ between the initial and final
tetraquarks and factoring out the lepton fields we get the matrix element of
the current%
\begin{equation}
J_{\mu }^{\mathrm{tr}}=\overline{c}\gamma _{\mu }(1-\gamma _{5})b
\label{eq:TrCurr}
\end{equation}%
in terms of the form factors $G_{i}(q^{2})$ that parametrize the
long-distance dynamics of the weak transition \cite{Ball:1991bs}
\begin{eqnarray}
&&\langle Z(p^{\prime })|J_{\mu }^{\mathrm{tr}}|T(p,\epsilon )\rangle =%
\widetilde{G}_{0}(q^{2})\epsilon _{\mu }+\widetilde{G}_{1}(q^{2})(\epsilon
p^{\prime })P_{\mu }  \notag \\
&&+\widetilde{G}_{2}(q^{2})(\epsilon p^{\prime })q_{\mu }+i\widetilde{G}%
_{3}(q^{2})\varepsilon _{\mu \nu \alpha \beta }\epsilon ^{\nu }p^{\alpha
}p^{\prime }{}^{\beta }.  \label{eq:Vertex}
\end{eqnarray}%
The scaled functions $\widetilde{G}_{i}(q^{2})$ above are connected with the
dimensionless form factors $G_{i}(q^{2})$ by the following equalities
\begin{equation}
\widetilde{G}_{0}(q^{2})=\widetilde{m}G_{0}(q^{2}),\ \widetilde{G}%
_{j}(q^{2})=\frac{G_{j}(q^{2})}{\widetilde{m}},\ j=1,2,3.  \label{eq:VertexA}
\end{equation}%
In Eqs.\ (\ref{eq:Vertex}) and \ (\ref{eq:VertexA}) $\widetilde{m}=m+m_{Z},\
p$ and $\epsilon $ are the momentum and polarization vector of the
tetraquark $T_{bb;\overline{u}\overline{d}}^{-}$, $p^{\prime }$ is the
momentum of the state $Z_{bc}^{0}$, $P_{\mu }=p_{\mu }^{\prime }+p_{\mu }$,
and $q_{\mu }=p_{\mu }-p_{\mu }^{\prime }$ is the momentum transferred to
the leptons. It is clear that $q^{2}$ changes within the limits $%
m_{l}^{2}\leq q^{2}\leq (m-m_{Z})^{2},$ where $m_{l}$ is the mass of the
lepton $l$.

The form factors $G_{i}(q^{2})$ are quantities that should be extracted
from the sum rules which, in turn, are obtainable from an analysis of the
three-point correlation function
\begin{eqnarray}
\Pi _{\mu \nu }(p,p^{\prime }) &=&i^{2}\int d^{4}xd^{4}ye^{i(p^{\prime
}y-px)}  \notag \\
&&\times \langle 0|\mathcal{T}\{J^{Z}(y)J_{\mu }^{\mathrm{tr}}(0)J_{\nu
}^{^{\dagger }}(x)\}|0\rangle ,  \label{eq:CF7}
\end{eqnarray}%
where $J_{\nu }(x)$ and $J^{Z}(y)$ are the interpolating currents to the $%
T_{bb;\overline{u}\overline{d}}^{-}$ and $Z_{bc}^{0}$ $\ $states,
respectively.

To derive sum rules for the weak form factors we express the correlation
function $\Pi _{\mu \nu }(p,p^{\prime })$ in terms of the masses and
couplings of the involved particles, and thus determine the
physical or phenomenological side of the sum rule $\Pi _{\mu \nu }^{\mathrm{%
Phys}}(p,p^{\prime })$. We also calculate $\Pi _{\mu \nu }(p,p^{\prime })$
using the interpolating currents and quark propagators, which leads to its
expression in terms of the quark, gluon, and mixed vacuum condensates. By
matching the obtained results and employing the assumption on the
quark-hadron duality, it is possible to extract sum rules and evaluate the
physical parameters of interest.

The function $\Pi _{\mu \nu }^{\mathrm{Phys}}(p,p^{\prime })$ can be easily
written down in the form%
\begin{eqnarray}
&&\Pi _{\mu \nu }^{\mathrm{Phys}}(p,p^{\prime })=\frac{\langle
0|J^{Z}|Z(p^{\prime })\rangle \langle Z(p^{\prime })|J_{\mu }^{\mathrm{tr}%
}|T(p,\epsilon )\rangle }{(p^{2}-m^{2})(p^{\prime 2}-m_{Z}^{2})}  \notag \\
&&\times \langle T(p,\epsilon )|J_{\nu }^{^{\dagger }}|0\rangle +\ldots ,
\label{eq:Phys1}
\end{eqnarray}%
where we only take into account contribution arising from the ground-state
particles, and effects of the excited and continuum
states are denoted by dots.

The phenomenological side of the sum rules can be further simplified by
rewriting the relevant matrix elements in terms of the tetraquarks
parameters, and employing for $\langle Z(p^{\prime })|J_{\mu }^{\mathrm{tr}%
}|T(p,\epsilon )\rangle $ its expression through the weak transition form
factors $G_{i}(q^{2})$. The matrix elements of the tetraquarks $T_{bb;%
\overline{u}\overline{d}}^{-}$ and $Z_{bc}^{0}$ are known and given by Eqs.\
(\ref{eq:MElem1}) and (\ref{eq:MElem2}), respectively. The matrix element $%
\langle Z(p^{\prime })|J_{\mu }^{\mathrm{tr}}|T(p,\epsilon )\rangle $ is
modeled by means of the four transition form factors $G_{i}(q^{2})$ which
can be used to calculate all  three semileptonic decays.

Substituting the relevant matrix elements into Eq.\ (\ref{eq:Phys1}), for $%
\Pi _{\mu \nu }^{\mathrm{Phys}}(p,p^{\prime },q^{2})$ we finally get
\begin{eqnarray}
&&\Pi _{\mu \nu }^{\mathrm{Phys}}(p,p^{\prime },q^{2})=\frac{fmf_{Z}m_{Z}}{%
(p^{2}-m^{2})(p^{\prime 2}-m_{Z}^{2})}  \notag \\
&&\times \left\{ \widetilde{G}_{0}(q^{2})\left( -g_{\mu \nu }+\frac{p_{\mu
}p_{\nu }}{m^{2}}\right) +\left[ \widetilde{G}_{1}(q^{2})P_{\mu }\right.
\right.  \notag \\
&&\left. +\widetilde{G}_{2}(q^{2})q_{\mu }\right] \left( -p_{\nu }^{\prime }+%
\frac{m^{2}+m_{Z}^{2}-q^{2}}{2m^{2}}p_{\nu }\right)  \notag \\
&&\left. -i\widetilde{G}_{3}(q^{2})\varepsilon _{\mu \nu \alpha \beta
}p^{\alpha }p^{\prime }{}^{\beta }\right\} +\ldots   \label{eq:Phys2}
\end{eqnarray}

The function $\Pi _{\mu \nu }^{\mathrm{OPE}}(p,p^{\prime })$ constitutes the
second side of the sum rules and has the following form
\begin{eqnarray}  \label{eq:CF8}
&&\Pi _{\mu \nu }^{\mathrm{OPE}}(p,p^{\prime })=\int
d^{4}xd^{4}ye^{i(p^{\prime }y-px)}\left\{ \mathrm{Tr}\left[ \gamma _{5}%
\widetilde{S}_{d}^{b^{\prime }b}(x-y)\right. \right.  \notag \\
&&\left. \times \gamma _{5}S_{u}^{a^{\prime }a}(x-y)\right] \left( \mathrm{Tr%
}\left[ \gamma _{\mu }\widetilde{S}_{b}^{aa^{\prime }}(y-x)\gamma
_{5}S_{c}^{bi}(y)\gamma _{\nu }(1-\gamma _{5})\right. \right.  \notag \\
&&\left. \times S_{b}^{ib^{\prime }}(-x)\right] +\mathrm{Tr}\left[ \gamma
_{\mu }\widetilde{S}_{b}^{ia^{\prime }}(-x)(1-\gamma _{5})\gamma _{\nu }%
\widetilde{S}_{c}^{bi}(y)\gamma _{5}\right.  \notag \\
&&\left. \left. \times S_{b}^{ab^{\prime }}(y-x)\right] \right) -\mathrm{Tr}%
\left[ \gamma _{5}\widetilde{S}_{d}^{b^{\prime }a}(x-y)\gamma
_{5}S_{u}^{a^{\prime }b}(x-y)\right]  \notag \\
&&\times \left( \mathrm{Tr}\left[ \gamma _{\mu }\widetilde{S}%
_{b}^{aa^{\prime }}(y-x)\gamma _{5}S_{c}^{bi}(y)\gamma _{\nu }(1-\gamma
_{5})S_{b}^{ib^{\prime }}(-x)\right] \right.  \notag \\
&&\left. \left. +\mathrm{Tr}\left[ \gamma _{\mu }\widetilde{S}%
_{b}^{ia^{\prime }}(-x)(1-\gamma _{5})\gamma _{\nu }\widetilde{S}%
_{c}^{bi}(y)\gamma _{5}S_{b}^{ab^{\prime }}(y-x)\right] \right) \right\}.
\notag \\
\end{eqnarray}%
To extract the sum rules for the form factors $G_{i}(q^{2})$ we equate
invariant amplitudes corresponding to the same Lorentz structures in $\Pi
_{\mu \nu }^{\mathrm{Phys}}(p,p^{\prime },q^{2})$ and $\Pi _{\mu \nu }^{%
\mathrm{OPE}}(p,p^{\prime })$, perform a double Borel transformation over
the variables $p^{\prime 2}$ and $p^{2}$ to suppress contributions of the higher
excited and continuum states, and perform continuum subtraction. For
example, to extract the sum rule for $\widetilde{G}_{0}(q^{2})$ we use the
structure $g_{\mu \nu }$, whereas for  $\widetilde{G}_{3}(q^{2})$ we employ the
term $\sim \varepsilon _{\mu \nu \alpha \beta }p^{\alpha }p^{\prime
}{}^{\beta }$. It is convenient to present the obtained sum rules in a single
formula through the functions $\widetilde{G}_{i}(q^{2})$,
\begin{eqnarray}
&&\widetilde{G}_{i}(\mathbf{M}^{2},\ \mathbf{s}_{0},~q^{2})=\frac{1}{%
fmf_{Z}m_{Z}}\int_{4m_{b}^{2}}^{s_{0}}ds\int_{(m_{b}+m_{c})^{2}}^{s_{0}^{%
\prime }}ds^{\prime }  \notag \\
&&\times \rho _{i}(s,s^{\prime
},q^{2})e^{(m^{2}-s)/M_{1}^{2}}e^{(m_{Z}^{2}-s^{\prime })/M_{2}^{2}},
\label{eq:FF}
\end{eqnarray}%
bearing in mind that they are connected to the dimensionless form factors $%
G_{i}(q^{2})$ by Eq.\ (\ref{eq:VertexA}). Here $\mathbf{M}^{2}=(M_{1}^{2},\
M_{2}^{2})$ are the Borel parameters, and $\mathbf{s}_{0}=(s_{0},\
s_{0}^{\prime })$ are the continuum threshold parameters that separate the
main contribution to the sum rules from the continuum effects. The sum rules
(\ref{eq:FF}) are written down using the spectral densities $\rho
_{i}(s,s^{\prime },q^{2})$ which are proportional to the imaginary parts of
the corresponding invariant amplitudes in $\Pi _{\mu \nu }^{\mathrm{OPE}%
}(p,p^{\prime })$. They contain the perturbative and nonperturbative
contributions, and are calculated with dimension-six accuracy.

For numerical computations of the weak form factors $G_{i}(\mathbf{M}^{2},\
\mathbf{s}_{0},~q^{2})$ one needs to fix various parameters. Values some of these
parameters are collected in  Table \ref{tab:Param}, while the masses and coupling
constants of the tetraquarks $T_{bb;\overline{u}\overline{d}}^{-}$ and $%
Z_{bc}^{0}$ were evaluated in the previous section. In the present
computations we impose the same constraints on the auxiliary parameters
$\mathbf{M}^{2}$ and $\mathbf{s}_{0}$  as in the mass calculations.

To obtain the width of the decay $T_{bb;\overline{u}\overline{d}%
}^{-}\rightarrow Z_{bc}^{0}l\overline{\nu }_{l}$ one has to integrate the
differential decay rate $d\Gamma /dq^{2}$ (for details, see the Appendix) within
allowed kinematical limits $m_{l}^{2}\leq q^{2}\leq (m-m_{Z})^{2}$. It is
clear that for light leptons $l=e,\ \mu $ the lower limit of the integral is
considerably smaller than $1\ \mathrm{GeV}^{2}$, but the perturbative
calculations lead to reliable predictions for momentum transfers $q^{2}>1\
\mathrm{GeV}^{2}$. Therefore, we use the usual prescription and replace the
weak form factors in the whole integration region by fit functions $%
F_{i}(q^{2})$ which for perturbatively allowed values of $q^{2}$ coincide
with $G_{i}(q^{2})$.

There are various analytical expressions for the fit functions. In the
present paper we utilize
\begin{equation}
F_{i}(q^{2})=f_{0}^{i}\exp \left[ c_{1i}\frac{q^{2}}{m_{\mathrm{fit}}^{2}}%
+c_{2i}\left( \frac{q^{2}}{m_{\mathrm{fit}}^{2}}\right) ^{2}\right] ,
\label{eq:FFunctions}
\end{equation}%
where $f_{0}^{i},~c_{1i},\ c_{2i}$ and $m_{\mathrm{fit}}^{2}$ are fitting
parameters. The values of these parameters are presented in Table \ref%
{tab:FitPar}. Besides that, for the numerical calculations we need the Fermi
coupling constant $G_{F}$ and CKM matrix element $|V_{bc}|$ for which we use
\begin{eqnarray}
G_{F} &=&1.16637\cdot 10^{-5}\ \mathrm{GeV}^{-2},\   \notag \\
|V_{bc}| &=&(41.2\pm 1.01)\cdot 10^{-3}.  \label{eq:Fermi}
\end{eqnarray}

\begin{table}[tbp]
\begin{tabular}{|c|c|c|c|c|}
\hline
$F_i(q^2)$ & $f_{0}^{i}$ & $c_{1i}$ & $c_{2i}$ & $m^2_{\mathrm{fit}}\ (%
\mathrm{GeV}^2)$ \\ \hline\hline
$F_{0}(q^2)$ & $-2.34$ & $19.53$ & $-36.87$ & $100.70$ \\ \hline
$F_{1}(q^2)$ & $-1.75$ & $18.45$ & $-14.29$ & $100.70$ \\ \hline
$F_{2}(q^2)$ & $8.80$ & $20.21$ & $-32.09$ & $100.70$ \\ \hline
$F_{3}(q^2)$ & $17.13$ & $20.60$ & $-32.49$ & $100.70$ \\ \hline
\end{tabular}%
\caption{The parameters of the fit functions $F_i(q^2)$.}
\label{tab:FitPar}
\end{table}

As a result, for the decay width of the processes $T_{bb;\overline{u}%
\overline{d}}^{-}\rightarrow Z_{bc}^{0}l\overline{\nu }_{l},\ (l=e,\ \mu $
and $\tau )$ we find%
\begin{eqnarray}
&&\Gamma \left( T_{bb;\overline{u}\overline{d}}^{-}\rightarrow Z_{bc}^{0}e%
\overline{\nu }_{e}\right) =\left( 2.65\pm 0.78\right) \cdot 10^{-8\ \ }%
\mathrm{MeV},  \notag \\
&&\Gamma \left( T_{bb;\overline{u}\overline{d}}^{-}\rightarrow Z_{bc}^{0}\mu
\overline{\nu }_{\mu }\right) =\left( 2.64\pm 0.78\right) \cdot 10^{-8\ \ }%
\mathrm{MeV},  \notag \\
&&\Gamma \left( T_{bb;\overline{u}\overline{d}}^{-}\rightarrow
Z_{bc}^{0}\tau \overline{\nu }_{\tau }\right) =\left( 1.88\pm 0.55\right)
\cdot 10^{-8\ }\mathrm{MeV},  \notag \\
&&  \label{eq:Width}
\end{eqnarray}%
which are the main results of the present work.

The partial decay widths from Eq.\ (\ref{eq:Width}) can be used to estimate
the full width and mean lifetime of the tetraquark $T_{bb;\overline{u}%
\overline{d}}^{-}$%
\begin{eqnarray}
\Gamma &=&(7.17\pm 1.23)\cdot 10^{-8\ \ }\mathrm{MeV},  \notag \\
\tau &=&9.18_{-1.34}^{+1.90}\cdot 10^{-15}\ \mathrm{s.}  \label{eq:WL}
\end{eqnarray}%
These predictions can be employed to explore the double-heavy tetraquarks.



\section{Analysis and conclusions}

\label{Analysis}
The spectroscopic parameters of the tetraquarks $T_{bb;\overline{u}\overline{d%
}}^{-}$ and $Z_{bc}^{0}$ as well as the width of the semileptonic decay $%
T_{bb;\overline{u}\overline{d}}^{-}\rightarrow Z_{bc}^{0}l\overline{\nu }%
_{l} $ provide very interesting information on the properties of four-quark
systems. Thus, the mass of the tetraquark $T_{bb;\overline{u}\overline{d}%
}^{-}$ obtained in the present work confirms once more that it is stable
against strong and electromagnetic decays, and can transform only weakly to
a tetraquark $Z_{bc}^{0}$ and a pair of leptons $l\overline{\nu }_{l}$. This
conclusion is valid even when  taking into account uncertainties inherent to the
sum rule computations. Our result for $m$ is smaller than the predictions
made in Refs.\ \cite{Navarra:2007yw} and \cite{Karliner:2017qjm} using the
QCD sum rule method and phenomenological model estimations, respectively.
The semileptonic decays $T_{bb;\overline{u}\overline{d}}^{-}\rightarrow
Z_{bc}^{0}l\overline{\nu }_{l}$, where $l=e,\ \mu $ and $\tau $ have allowed
us to evaluate the width of $T_{bb;\overline{u}\overline{d}}^{-}$ and its
mean lifetime $\tau =9.18_{-1.34}^{+1.90}~\mathrm{fs}$ which is considerably
shorter than the prediction of Ref.\ \cite{Karliner:2017qjm}.

Another interesting result of this work is connected with the parameters of the
scalar tetraquark $Z_{bc}^{0}$ composed of the heavy diquark $bc$ and light
antidiquark $\overline{u}\overline{d}$. In fact, the mass of this state $%
m_{Z}=(6660\pm 150~)\ \mathrm{MeV}$ is considerably below the threshold $%
\approx 7145\ \mathrm{MeV}$ for strong $S$-wave decays to conventional
heavy $B^{-}D^{+}$ and $\overline{B^{0}}D^{0}$ mesons. Because of its quark
content. $Z_{bc}^{0}$ cannot decay to a pair of heavy and light mesons as
well. These features differ  qualitatively from those of the open charm-bottom
scalar tetraquarks $Z_{q}=[cq][\overline{b}\overline{q}]$ and $Z_{s}=[cs][%
\overline{b}\overline{s}]$, which decay strongly to $B_{c}\pi $ and $%
B_{c}\eta $ mesons \cite{Agaev:2016dsg}, and, in turn, cannot decay to
two heavy mesons. In other words, the four-quark system consisting of a heavy
diquark and a light antidiquark is more stable than one consisting of a
heavy-light diquark and antidiquark. This is seen from a comparison of the
masses of the tetraquark $Z_{bc}^{0}$ and the state $Z_{q}$, for which $%
m_{Z_{q}}=(6.97\pm 0.19)~\mathrm{GeV}$.

Theoretical information on the decay properties of the state $T_{bb;%
\overline{u}\overline{d}}^{-}$ can be further improved by including
its other weak decay channels in analyses. The investigation of the stable open
charm-bottom tetraquarks $Z_{bc}^{0}$ with different quantum numbers is also
an interesting topic of exotic hadron physics:
by clarifying these
problems we can deepen our understanding of multiquark systems.


\section*{Acknowledgments}

S.~S.~A. is grateful to Prof. V.~M.~Braun for enlightening discussions.
K.~A., B.~B.,~ and H.~S.~ thank TUBITAK for the partial financial support
provided under Grant No. 115F183.

\appendix*

\section{ The decay rate $d\Gamma /dq^{2}$}

\renewcommand{\theequation}{\Alph{section}.\arabic{equation}} \label{sec:App}
This appendix contains the explicit expression for the decay rate $d\Gamma
/dq^{2}$ necessary to calculate the width of the semileptonic decay $T_{bb;%
\overline{u}\overline{d}}^{-}\rightarrow Z_{bc}^{0}l\overline{\nu }_{l}$.
Calculations lead to the following result:
\begin{widetext}

\begin{eqnarray}
&&\frac{d\Gamma }{dq^{2}}=\frac{G_{F}^{2}|V_{cb}|^{2}}{3\cdot 2^{8}\pi
^{3}m^{3}}\left( \frac{q^{2}-m_{l}^{2}}{q^{2}}\right) \lambda \left(
m^{2},m_{Z}^{2},q^{2}\right) \left[ \sum_{i=0}^{i=3}\widetilde{G}%
_{i}^{2}(q^{2})\mathcal{A}_{i}(q^{2})+\widetilde{G}_{0}(q^{2})\widetilde{G}%
_{1}(q^{2})\mathcal{A}_{01}(q^{2})\right.  \notag \\
&&\left. +\widetilde{G}_{0}(q^{2})\widetilde{G}_{2}(q^{2})\mathcal{A}%
_{02}(q^{2})+\widetilde{G}_{1}(q^{2})\widetilde{G}_{2}(q^{2})\mathcal{A}%
_{12}(q^{2})\right. \bigg],  \label{eq:DR}
\end{eqnarray}%
In Eq.\ (\ref{eq:DR}) the functions $\mathcal{A}_{i}(q^{2})$ and $\mathcal{A}%
_{ij}(q^{2})$ are given by

\begin{eqnarray}
\mathcal{A}_{0}(q^{2}) &=&\frac{1}{2m^{2}q^{4}}\left[ q^{4}\left(
m^{2}-m_{Z}^{2}\right) ^{2}-4q^{4}m^{2}m_{l}^{2}-m_{l}^{4}\left(
m^{2}-m_{Z}^{2}+q^{2}\right) ^{2}+2q^{6}\left( 3m^{2}-m_{Z}^{2}\right) +q^{8}%
\right] ,  \notag \\
\mathcal{A}_{1}(q^{2}) &=&\frac{1}{2m^{2}q^{4}}\left[ m^{4}+\left(
m_{Z}^{2}-q^{2}\right) ^{2}-2m^{2}(m_{Z}^{2}+q^{2})\right] \left\{
m_{l}^{4}(m^{2}-m_{Z}^{2})^{2}+q^{4}m_{l}^{4}(q^{2}-2m^{2}-2m_{Z}^{2})\right.
\notag \\
&&\left. -q^{4}\left[ m^{4}+(m_{Z}^{2}-q^{2})^{2}-2m^{2}(m_{Z}^{2}+q^{2})%
\right] \right\} ,  \notag \\
\mathcal{A}_{2}(q^{2}) &=&\frac{m_{l}^{2}}{2m^{2}}\left(
q^{2}-m_{l}^{2}\right) \left[
m^{4}+(m_{Z}^{2}-q^{2})^{2}-2m^{2}(m_{Z}^{2}+q^{2})\right] ,  \notag \\
\mathcal{A}_{3}(q^{2}) &=&\frac{1}{2q^{2}}(m_{l}^{4}-q^{4})\left[
m^{4}+(m_{Z}^{2}-q^{2})^{2}-2m^{2}(m_{Z}^{2}+q^{2})\right] ,  \notag \\
\mathcal{A}_{01}(q^{2}) &=&\frac{1}{m^{2}q^{4}}\left[
q^{4}(m_{l}^{2}+m_{Z}^{2}-m^{2}-q^{2})+m_{l}^{4}(m^{2}-m_{Z}^{2})\right] %
\left[ m^{4}+(m_{Z}^{2}-q^{2})^{2}-2m^{2}(m_{Z}^{2}+q^{2})\right] ,  \notag
\\
\mathcal{A}_{02}(q^{2}) &=&\frac{m_{l}^{2}(m_{l}^{2}-q^{2})}{m^{2}q^{2}}%
\left[ m^{4}+(m_{Z}^{2}-q^{2})^{2}-2m^{2}(m_{Z}^{2}+q^{2})\right] ,  \notag
\\
\mathcal{A}_{12}(q^{2}) &=&\frac{m_{l}^{2}(q^{2}-m_{l}^{2})(m^{2}-m_{Z}^{2})%
}{m^{2}q^{2}}\left[ m^{4}+(m_{Z}^{2}-q^{2})^{2}-2m^{2}(m_{Z}^{2}+q^{2})%
\right] ,  \label{eq:Func}
\end{eqnarray}%
and%
\begin{equation*}
\lambda \left( m^{2},m_{Z}^{2},q^{2}\right) =\left[ m^{4}+m_{Z}^{4}+q^{4}-2%
\left( m^{2}m_{Z}^{2}+m^{2}q^{2}+m_{Z}^{2}q^{2}\right) \right] ^{1/2}.
\end{equation*}

\end{widetext}

\end{document}